\documentclass[12pt,a4paper]{article}
\usepackage{graphicx}

\title
{ The Reaction $D(e,pp)e'\pi^-$ on Polarized
Deuteron at High Proton Momenta}

\author{ 
V.N. Stibunov$^{1}$, A.Yu. Loginov$^{1}$, D.M. Nikolenko$^{2}$, A.V. Osipov$^{1}$,\\
I.A. Rachek$^{2}$,  A.A. Sidorov$^{1}$,  D.K. Toporkov$^{2}$\\
\and
\it $^1$Nuclear Physics Institute at Tomsk Polytechnical University, \\
\it Tomsk, Russia
\and
\it $^2$Budker Institute of Nuclear Physics,  \\
\it Novosibirsk, Russia
}
\begin{document}
\date{}
\maketitle

\begin{center}
\large
{Invited talk presented at the 16th European Conference\\
 on Few-Body Problems in Physics, \\
Autrans, France, June 1-6, 1998}
\\[5mm]
\end{center}

\sloppy

\begin{abstract}
The differential cross section and target asymmetry components 
of the reaction $D(e,pp)e'\pi^-$ on polarized deuteron were measured. 
The kinetic energies of the protons were measured within 55-180 MeV and
 46-265 MeV and the acceptance angles in lab. frame are $\Theta_{1,2}=64^0 - 82^0$, 
$\Delta\phi_{1,2} = 32^0$. The sharp peak of the tensor $a_{20}$-component of the target asymmetry is found near the invariant mass
of the $pp\pi$-system $M_{pp\pi}$ = 2300 $MeV/c^2$. The performed calculations of the differential yield and  the tensor target asymmetry do not describe the
obtained experimental results.
\end{abstract}

The interest to 
study  of the  $\pi^-$-meson production on  a deuteron
for high polar angles and large momenta of both protons proceeds  from
an  opportunity  to acquire a
new information on the dynamics of NN-interaction at short  internucleon distances. In the region of proton
momenta  larger than the Fermi-momentum the quasifree mechanism of the $\pi^-$-meson production appears to be suppressed.  The relative contribution   of  more
complex reaction mechanisms grows in this kinematic area
and these  reactions require
new models to describe the nucleon systems and hadron interactions. It
is for these reasons  the previous experiments  in Hamburg \cite{B1} , 
Saclay \cite{A2}  and Bonn  \cite{B3,R4} chose the search for dibarion states and    observation  of the $(\Delta\Delta)$-states as their main subject.

   Our experiment was focused on the region of an even  higher  opening
angles and larger values of the invariant mass than before. Also, the use
of a polarized deuteron target enabled us to consider a number of  
polarization observables.
  
   The measurements reported here were  conducted  simultaneously  with
the  experiments  performed   \cite{G5,M6}, 
which used an internal
tensor-polarized  deuterium target in the VEPP-3  storage ring
at 2 GeV electron energy.The particle-detection system 
consisted of two identical two-arm apparatus to detect the protons in coincidence  \cite{I7,L8}. Each arm of the detector was placed symmetrically around the
electron-beam axis at a polar angle of $75^0$  with respect  to  the
beam line.  The proton telescope included a drift chamber and the thin
and thick scintillator counters.  Each proton arm  detected  particles
within  the  range  of  angles  $\theta  = 68^o$ -- 82$^o$ and $\Delta
\varphi = 32^o$.  The kinetic energy of the protons which  deposit  all
their  energy was reconstructed combining the values of the energy 
deposition in the detector layers.   In these measurements the direction and sign  of  the
target  polarization were changed periodically during the data 
acquisition \cite{M6}.  
The integrated luminosity and the average value of  the  tensor
target  polarization  were  determined  from electron-deuteron elastic
scattering \cite{G5}.
 
  The collected data   were processed in a few 
consecutive stages \cite{L8,O9},
which resulted in momentum vectors for both protons and  reconstructed
the vertex coordinates for the events.  Computation of the pion momentum
and photon energy was done on an assumption  of  a  zero  angle  of
electron scattering.
  The selected events were used to determine the yield of the  reaction  to  each  detector for two signs and two directions of the
guide magnetic field. 
   
The components of the experimental target  asymmetry are defined as the  counting rate combinations  \cite{M6} :
\begin{eqnarray}
  a_{11} = 
\frac{\sum_{i=1,2}
(-1)^{\delta_{i1}}\left[N_{1+}^i + N_{2-}^i\right]}{N}, \ \ \ \
a_{20}  =  \frac{\sum_{i,j=1,2}(
N_{j+}^i -  N_{j-}^i)}{N} , \nonumber\\
a_{22} =  
\frac{\sum_{i,j=1,2}( - 1)^{\delta_{ij}}[ N_{j-}^i - N_{j+}^i]}{N},
\end{eqnarray}
\medskip

 where $N_{jk}^i$ is the counting rate in the detector system $i$ with
the magnetic guide field index $j$, and the sign of deuteron tenzor polarization degree $P_{zz}$ given by $k$
and N is the total counting rate.

In order to  obtain the distributions of
   We used  (see . ref. \cite{L8}) the connection between  the 
yield of the reaction summed over $i$, $j$ and $k$ into a 6-D phase space volume of the momenta, $V_6$ and differential cross  section of the reaction

\begin{eqnarray}
Y(V_6)=\int_{V_6}\epsilon L
\frac{d^6\sigma}{d^3p_{1}d^3p_{2}}d^3p_1d^3p_2 ,
\end{eqnarray}

where $p_{1}$ and $p_{2}$ are the momenta  of the protons, $\epsilon$ is the  total detection and selection efficiency  of the $pp$- events,   $L$ is
the integral luminosity obtained from the measured elastic $ed$- scattering. 

   The dependences of the cross section and the analyzing power components of the reaction on the invariant mass $pp$-system that we  obtaned at this experiment were presented in  ref. \cite{L8,O9}. Here we  present the first results as a function of  the $pp\pi^-$-system  mass, $M_{pp\pi}$. The differential yield of the reaction is shown in Fig.1. and tensor $a_{20}$-component of the target asymmetry is shown in Fig.2. 
 
    The calculations of the cross section of the 
investigated process were made in a few theoretical models. The cross
section  of the process initiated by electrons was expressed in the terms of the cross section of a reaction induced by the virtual photons.  We used 
Dalitz-Yennie's   virtual photon spectrum. For NEWGAM-code \cite{L8} one  nucleon pion  photoproduction  operator  has been taken from the phenomenological  analysis \cite{M10} and the deuteron wave function was
obtained using the Paris N-N potential.  Also we used the ENIGMA-code which was developed  for the exclusive pion electroproduction on nuclei \cite{V11}. 
The  calculations of the polarization observables				        and cross section of the reaction were made within the spectator model using the elementary pion  photoproduction amplitude discussed in ref. \cite{D12}. 
The Born terms of this amplitude are determined in pseudovector $\pi {N}$-coupling, the $\Delta$-resonance is considered  both in the $s$- and the $u$-channels and the $\rho$ and $\omega$-mesons exchange are considered in the $t$-channel. This
amplitude is useful for the studies of  the  $\Delta$-resonance. In addition,
we studied the role of the various dynamic effects in photoproduction of $\Delta(1232)$-isobar on a polarized deuteron using the relativistic impulse approximation in the nucleon-spectator model \cite{L13}. 
\begin{figure}[ht]
\includegraphics[width=0.47\textwidth]{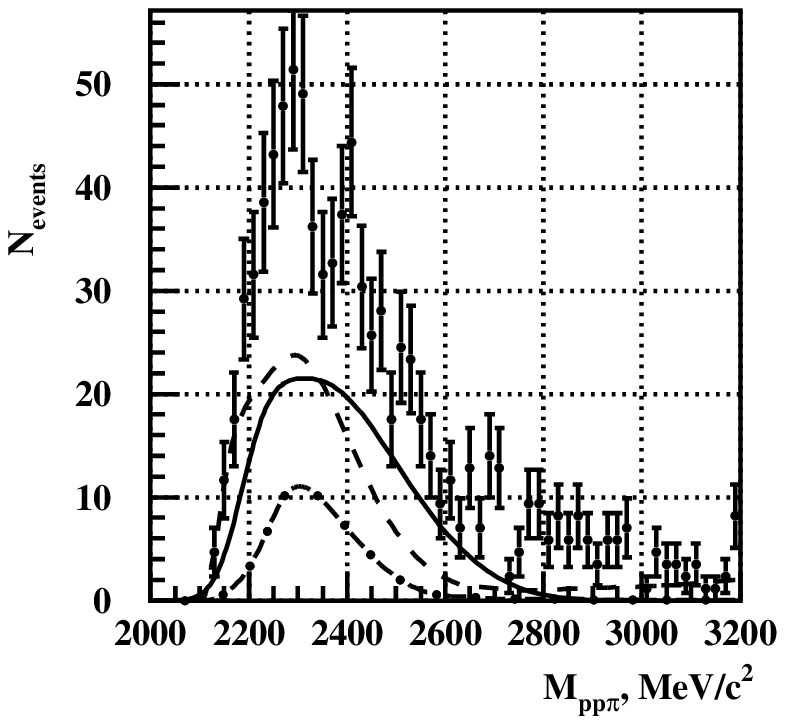}
\hfill
\includegraphics[width=0.47\textwidth]{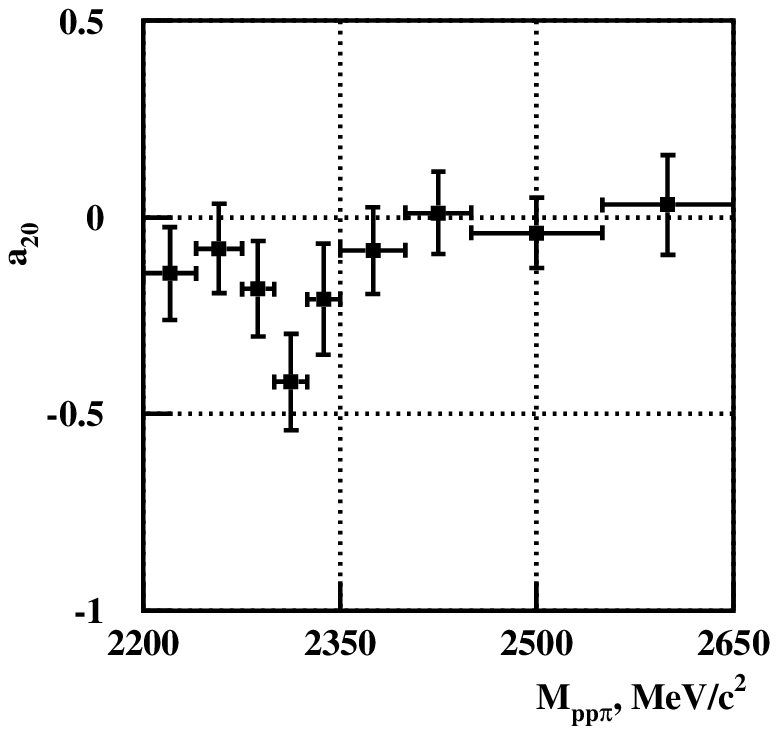}
\parbox[t]{0.47\textwidth}{
\caption{
Differential yield of the reaction 
$d(e,pp)e'\pi^-$ depending on the $pp\pi^-$-- invariant mass.
Solid line -- ENIGMA. Dashed line -- full one-nucleon amplitude. Dot-dashed 
line -- one-nucleon amplitude with $\Delta$-pole in  $s$-channel only.
}\label{energ}}
\hfill
\parbox[t]{0.47\textwidth}{
\caption{
Tensor  $a_{20}$ -- component of the target asymmetry
as a function of the $pp\pi^-$ invariant mass.
}\label{mpp}}
\end{figure}

  The experimental and some calculated dependences of the
differential yield of the reaction
 on the mass of the $pp\pi$-system can be see in Fig. 1.The  solid  curve 
shows the result of the ENIGMA-code, the result of the NEWGAM-code is slightly different from this result. The dashed line
corresponds to the calculation based on the  total one-nucleon amplitude of the $\pi$-mesons photoproduction \cite{D12}, whereas the dot-dashed line shows the result of the calculation based on the  one-nucleon amplitude  including  
the $\Delta$-isobar in the $s$-channel only. One can see that the experimental
spectrum  is peaked at $M_{pp\pi}$ = 2300 $MeV/c^2$. It is clear from 
this figure that
the experimental yield of the reaction are much higher than their  calculated
counterparts.

   Figure 2. plots the behavior 
of the tensor target  $a_{20}$ --asymmetry versus the invariant mass of the $pp\pi^-$-system. Here one can see a peculiar feature - an 
sharp rise in the range of masses
$M_{pp\pi}$=2300  $MeV/c^2$. Note that events from this range correspond to production $\Delta^{0}(1232)$-isobar.  This could be seen  from the distribution of the invariant mass pion and one  the fastest of two protons, $M_{p\pi}$  - which excibite a clean peak at 1232 $MeV/c^2$. The calculated values of
the $a_{20}$-component of the target asymmetry is below 0.6 in the mass region near 2300 $MeV/c^2$. 

We keep on working on further analysis of obtained results.
These results allow us to make two conclusions. The behavior of the differential yield and $a_{20}$ -component of the target asymmetry near $M_{pp\pi}$=2300 $MeV/c^2$ is associated with  
the excitation  of $\Delta^{0}(1232)$-isobar on the deuteron.
  The noticeable difference 
 near $M_{pp\pi}$=2300 $MeV/c^2$ between the experimental values and the 
 calculated  tensor target  $a_{20}$  asymmetry and the reaction yield may 
be related with an excitation of a dibarion  resonance state and  its following
decay into proton and $\Delta^{0}(1232)$-isobar.

    This work was supported by  the Russian Foundation  for Fundamental Research grants No.98-02-17993, No.98-02-17949 and by  the INTAS grant No.96-0424.

\end{document}